\documentclass[twocolumn,showpacs,prc,superscriptaddress,floatfix,amsmath,amssymb]{revtex4}
\usepackage{graphicx}% Include figure files
\usepackage{dcolumn}% Align table columns on decimal point
\usepackage{bm}% bold math
\usepackage[usenames]{color}

\begin{document}

\preprint{APS/123-QED}

%\title{Penta-quark search via ($\pi^{-}, K^{-}$) reaction}
\title{Search for $\Theta^{+}$ via $K^{+}p \to  \pi^{+}X$ reaction with a 1.2 GeV/$c$ $K^{+}$ beam}

\author{K.~Miwa}
\altaffiliation[Corresponding author, email: miwa9@lambda.phys.tohoku.ac.jp, Present address: ]{Department of Physics, Tohoku University, Sendai 980-8578, Japan}
\affiliation{Department of Physics, Kyoto University, Kyoto 606-8502, Japan}

\author{S.~Dairaku}
\affiliation{Department of Physics, Kyoto University, Kyoto 606-8502, Japan}

\author{D.~Nakajima}
\affiliation{Department of Physics, University of Tokyo, 7-3-1 Hongo, Tokyo 113-0033, Japan}

\author{S.~Ajimura}
\altaffiliation[Present address: ]{Research Center for Nuclear Physics (RCNP), 10-1 Mihogaoka, Ibaraki, Osaka, 567-0047, Japan}
\affiliation{Department of Physics, Osaka University, Toyonaka 560-0043, Japan}

\author{J.~Arvieux}
\affiliation{Institut de Physique Nucl$\acute{\rm e}$aire, Universit$\acute{\rm e}$ Paris Sud, F-91406 Orsay cedex, France}

\author{ H.~Fujimura}
\affiliation{Department of Physics, Kyoto University, Kyoto 606-8502, Japan}

\author{H.~Fujioka}
\affiliation{Department of Physics, University of Tokyo, 7-3-1 Hongo, Tokyo 113-0033, Japan}

\author{T.~Fukuda}
\affiliation{Department of Physics, Osaka Electric Communication University, Osaka 558-8585, Japan}

\author{H.~Funahashi}
\altaffiliation[Present address: ]{Department of Physics, Osaka Electric Communication University, Osaka 558-8585, Japan}
\affiliation{Department of Physics, Kyoto University, Kyoto 606-8502, Japan}

\author{M.~Hayata}
\affiliation{Department of Physics, Kyoto University, Kyoto 606-8502, Japan}

\author{K.~Hicks}
\affiliation{Department of Physics, Ohio University, Athens, Ohio, USA 45701}

\author{K.~Imai}
\affiliation{Department of Physics, Kyoto University, Kyoto 606-8502, Japan}

\author{S.~Ishimoto}
\affiliation{KEK, High Energy Accelerator Research Organization, Tsukuba 305-0801, Japan}

\author{T.~Kameyama}
\affiliation{Physics Department, Gifu University, Gifu 501-1193, Japan}

\author{S.~Kamigaito}
\affiliation{Department of Physics, Kyoto University, Kyoto 606-8502, Japan}

\author{S.~Kinoshita}
\affiliation{Department of Physics, Tohoku University, Sendai 980-8578, Japan}

\author{T.~Koike}
\affiliation{Department of Physics, Tohoku University, Sendai 980-8578, Japan}

\author{Y.~Ma}
\affiliation{Department of Physics, Tohoku University, Sendai 980-8578, Japan}

\author{T.~Maruta}
\altaffiliation[Present address: ]{Department of Physics, Tohoku University, Sendai 980-8578, Japan}
\affiliation{Department of Physics, University of Tokyo, 7-3-1 Hongo, Tokyo 113-0033, Japan}

\author{Y.~Miura}
\affiliation{Department of Physics, Tohoku University, Sendai 980-8578, Japan}

\author{M.~Miyabe}
\affiliation{Department of Physics, Kyoto University, Kyoto 606-8502, Japan}

\author{T.~Nagae}
\altaffiliation[Present address: ]{Department of Physics, Kyoto University, Kyoto 606-8502, Japan}
\affiliation{KEK, High Energy Accelerator Research Organization, Tsukuba 305-0801, Japan}

\author{T.~Nakano}
\affiliation{Research Center for Nuclear Physics (RCNP), 10-1 Mihogaoka, Ibaraki, Osaka, 567-0047, Japan}

\author{K.~Nakazawa}
\affiliation{Physics Department, Gifu University, Gifu 501-1193, Japan}

\author{M.~Naruki}
\altaffiliation[Present address: ]{KEK, High Energy Accelerator Research Organization, Tsukuba 305-0801, Japan}
\affiliation{RIKEN, 2-1 Hirosawa, Wako, Saitama 351-0198, Japan}

\author{H.~Noumi}
\altaffiliation[Present address: ]{Research Center for Nuclear Physics (RCNP), 10-1 Mihogaoka, Ibaraki, Osaka, 567-0047, Japan}
\affiliation{KEK, High Energy Accelerator Research Organization, Tsukuba 305-0801, Japan}

\author{M.~Niiyama}
\altaffiliation[Present address: ]{RIKEN, 2-1 Hirosawa, Wako, Saitama 351-0198, Japan}
\affiliation{Department of Physics, Kyoto University, Kyoto 606-8502, Japan}

\author{N.~Saito}
\altaffiliation[Present address: ]{Present address: KEK, High Energy Accelerator Research Organization, Tsukuba 305-0801, Japan}
\affiliation{Department of Physics, Kyoto University, Kyoto 606-8502, Japan}

\author{Y.~Sato}
\affiliation{KEK, High Energy Accelerator Research Organization, Tsukuba 305-0801, Japan}

\author{S.~Sawada}
\affiliation{KEK, High Energy Accelerator Research Organization, Tsukuba 305-0801, Japan}

\author{Y.~Seki}
\affiliation{Department of Physics, Kyoto University, Kyoto 606-8502, Japan}

\author{M.~Sekimoto}
\affiliation{KEK, High Energy Accelerator Research Organization, Tsukuba 305-0801, Japan}

\author{K.~Senzaka}
\affiliation{Department of Physics, Kyoto University, Kyoto 606-8502, Japan}

\author{K.~Shirotori}
\affiliation{Department of Physics, Tohoku University, Sendai 980-8578, Japan}

\author{K.~Shoji}
\affiliation{Department of Physics, Kyoto University, Kyoto 606-8502, Japan}

\author{S.~Suzuki}
\affiliation{KEK, High Energy Accelerator Research Organization, Tsukuba 305-0801, Japan}

\author{H.~Takahashi}
\affiliation{KEK, High Energy Accelerator Research Organization, Tsukuba 305-0801, Japan}

\author{T.~Takahashi}
\affiliation{KEK, High Energy Accelerator Research Organization, Tsukuba 305-0801, Japan}
\author{T.N.~Takahashi}
\affiliation{Department of Physics, University of Tokyo, 7-3-1 Hongo, Tokyo 113-0033, Japan}

\author{H.~Tamura}
\affiliation{Department of Physics, Tohoku University, Sendai 980-8578, Japan}

\author{N.~Tanaka}
\affiliation{KEK, High Energy Accelerator Research Organization, Tsukuba 305-0801, Japan}

\author{K.~Tanida}
\altaffiliation[Present address: ]{Department of Physics, Kyoto University, Kyoto 606-8502, Japan}
\affiliation{RIKEN, 2-1 Hirosawa, Wako, Saitama 351-0198, Japan}

\author{A.~Toyada}
\affiliation{KEK, High Energy Accelerator Research Organization, Tsukuba 305-0801, Japan}

\author{T.~Watanabe}
\affiliation{Physics Department, Gifu University, Gifu 501-1193, Japan}

\author{M.~Yosoi}
\affiliation{Research Center for Nuclear Physics (RCNP), 10-1 Mihogaoka, Ibaraki, Osaka, 567-0047, Japan}

\author{R.~Zavislak}
\affiliation{Department of Physics, Ohio University, Athens, Ohio, USA 45701}

\date{\today}% It is always \today, today,

\begin{abstract}
The $\Theta^{+}$ was searched for via the $K^{+}p \to  \pi^{+}X$ reaction 
using the 1.2 GeV/$c$ $K^{+}$ beam at the K6 beam line of 
the KEK-PS 12 GeV Proton Synchrotron.
In the missing mass spectrum of the $K^{+}p \to \pi^{+}X$ reaction,
no clear peak structure was observed.
Therefore a 90 \% C.L. upper limit of 3.5 $\mu$b/sr
was derived for the differential cross section averaged 
over 2$^{\rm o}$ to 22$^{\rm o}$ in the laboratory frame
of the $K^{+}p \to \pi^{+}\Theta^{+}$ reaction. 
This upper limit is much smaller than the theoretical calculation
for the $t$-channel process where a $K^{0*}$ is exchanged.
From the present result, either the $t$-channel process is excluded or the
coupling constant of g$_{K*N\Theta}$ is quite small.
\end{abstract}
\maketitle

\section{Introduction}
Since the first report giving evidence of the existence of an 
exotic baryon $\Theta^{+}$ \cite{LEPS}, 
many papers from 
both theoretical and experimental aspects have been published 
\cite{review, Goeke}.
The $\Theta^{+}$ was observed as a narrow resonance of a $K^{+}n$ system, 
giving its minimum quark content as $uudd\bar{s}$. 
Therefore, if the $\Theta^{+}$ would exist, it could have interesting 
consequences for non-perturbative QCD.
The first observation by the LEPS collaboration was confirmed by 
several experiments 
\cite{DIANA, CLAS, SAPHIR, CLAS2, nu, HERMES, ZEUS, COSY, SVD, C3H8}.
However, null results were also reported from several high energy 
experiments where they searched for the $\Theta^{+}$ with much higher 
statistics \cite{HyperCP, HERA-B, ALEPH, BES, BABAR, CDF, SPHINX}.
Moreover, some of the initial positive evidence was refuted by the same 
collaboration with higher statistics \cite{CLAS3, CLAS4, COSY2}.
On the other hand, the LEPS and DIANA collaborations have shown
new evidence supporting its existence.
The LEPS collaboration shows a narrow peak in the missing mass spectrum
of the $\gamma d \to \Lambda(1520) X$ reaction \cite{LEPS2}.
The DIANA collaboration confirmed its initial positive evidence with two
times larger statistics \cite{DIANA2}.
In this controversial situation, 
it is quite important to perform 
other experimental searches that 
would reveal the production mechanism of the
$\Theta^{+}$ by using various reactions.

If the $\Theta^{+}$ exists, one of its most remarkable features would be
its narrow width.
The observed width is consistent with the experimental resolution in all 
positive experiments.
Moreover reanalyses of the past $K^{+}$-nucleon elastic scattering 
restricted the width to be less than a few MeV/$c^{2}$ \cite{Arndt}.
Cahn and Trilling 
calculated the width from the result of the DIANA experiment,
where the $\Theta^{+}$ was observed via the charge exchange channel of
$K^{+}n \to K^{0}p$, and obtained the width of 0.9 $\pm$ 0.3 MeV/$c^{2}$
\cite{Cahn}.
Such narrow width is unusual for a hadron resonance.
%If the width of the $\Theta^{+}$ is really narrow, it should be related to
%the internal dynamics of quarks.
In order to understand the narrow width, theoretical models were proposed
that take into account an additional correlation between quarks 
\cite{Diakonov, JaffeWilczek, Lipkin, Kishimoto, Enyo, Hiyama}.
The different dynamics of several models results in 
different values of the spin and parity.
Therefore experimentally, if the $\Theta^+$ exists,
it is quite important to determine its spin and parity to understand the
nature of the $\Theta^{+}$ and hence its quark dynamics.
For this purpose, experiments with high statistics are needed. 
Hadronic reactions fulfill these requirements,
because the production cross section via a hadronic reaction is
expected to be much larger than that of a photoproduction reaction.
Theoretically, the cross section of the $K^{+}p \to \pi^{+}\Theta^{+}$
reaction is expected to be $\sim$80$\mu$b \cite{Oh2}. 

High-resolution spectroscopy is another important point.
Because the width of the $\Theta^{+}$ is expected to be narrow,
high resolution is essential in order to measure its width 
with high experimental sensitivity.
Therefore we were motivated to do an experiment using 
a $K^{+}$ beam and a high resolution spectrometer to search for 
the $\Theta^{+}$ with high statistics.

%
%The KEK-PS E559 experiment was performed to search for the $\Theta^{+}$ via 
%$K^{+}p \to \pi^{+}X$ reaction using the SKS spectrometer at the K6 beam line
%at KEK 12GeV Proton Synchrotron.
%The present experiment has two advantages.
%One thing is the excellent experimental resolution of 2.4 MeV/$c^{2}$ (FWHM)
%which is consistent with the expected width of the $\Theta^{+}$.
%In this reaction,
%there are large background events which originate from
%$\Delta$ and $K^{*}$ productions whose cross section is $\sim$5mb \cite{Bland},
%in addition to decay events of $K^{+}$ beams.
%If the resolution is poor, the sensitivity becomes quite low.
%The SKS provides a high mass resolution leading a high sensitivity.
%Therefore this experiment is unique one.
%The other advantage is that the production cross section is expected to
%be large because we use $K^{+}$ beams including $\bar{s}$ quark. 

A different experiment,
the E522 collaboration, reported a search for the $\Theta^{+}$
via the $\pi^{-} p \to K^{-} X$ reaction \cite{e522}, finding 
a bump structure near 1.53 GeV/$c^{2}$ 
in the missing mass spectrum off the $K^-$.
The statistical significance of the bump was 2.5~-~2.7$\sigma$ 
and was not sufficient 
to claim the existence of the $\Theta^{+}$.
The upper limit of the production cross section was found to be 3.9 $\mu$b.
This reaction is just the inverse reaction of $K^{+} p \to \pi^{+} X$.
In both reactions, 
the same coupling constants are used in the production diagrams.
%Naively the production cross section via $K^{+} p \to \pi^{+} \Theta^{+}$
%reaction is expected to be larger than that via 
%$\pi^{-} p \to K^{-} \Theta^{+}$ reaction, because $K^{+}$ beam particles
%already include $\bar{s}$ quarks.
A deeper understanding on the production mechanism can be obtained
by considering both experimental results.

Theoretical calculations in hadronic models using effective 
interaction Lagrangians and form factors were made 
by several authors \cite{Ko, Oh, Oh2, Hyodo, Hyodo2}.
They studied the $\Theta^{+}$ production mechanism in
$\gamma N, NN, KN$ and $\pi N$ reactions near the production threshold 
comprehensively.
In general all calculations predict a large production cross section
for the $K^{+}p \to \pi^{+}\Theta^{+}$ reaction.
For the production process, they took into account both 
the t-channel process, where $K^{0*}$ was exchanged, and 
the u-channel process, which includes intermediate $N^{*}$ states.
%The contribution of t-channel is significant for the reaction
%and the angular distribution of the emitted $\pi^{+}$ has a forward peak.
%On the other hand, if the only u-channel process exists, the cross section 
%decreases and the angular distribution of the $\pi^{+}$ has a backward peak.
%Since the experimental spectrometer covers a forward angle 
%between 0 and 25 degrees in the laboratory frame, 
%the present experiment has a large sensitivity for the t-channel process.
In these calculations, unknown coupling constants, g$_{KN\Theta}$ and
g$_{K*N\Theta}$ were used.
Therefore measurements of both
$K^{+}p \to \pi^{+}\Theta^{+}$ and $\pi^{-}p \to K^{-}\Theta^{+}$ 
reactions are useful to compare with theory.

In this paper, 
we show the results of a search for the $\Theta^{+}$
via the $K^{+}p \to \pi^{+}X$ reaction.
We discuss the production mechanism considering both experimental results
of $K^{+}p \to \pi^{+}\Theta^{+}$ and $\pi^{-}p \to K^{-}\Theta^{+}$ reactions.

\section{Experiment}

\begin{figure*} [!t]
\includegraphics[width=12cm]{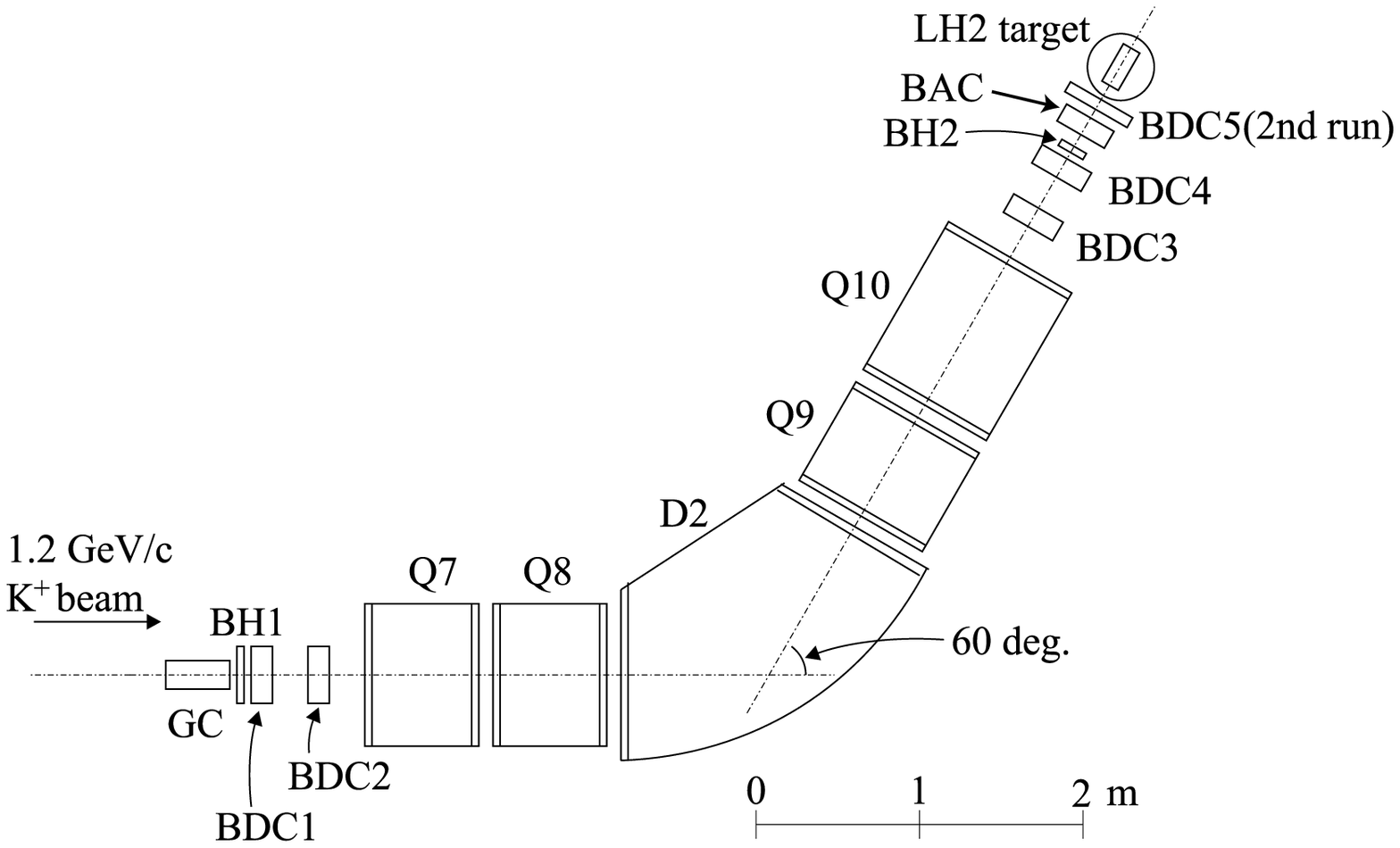}
\caption[]{Experimental setup of the K6 beam line spectrometer.
}
\label{k6_spectrometer}
\end{figure*}

\begin{figure*} [!t]
\includegraphics[width=12cm]{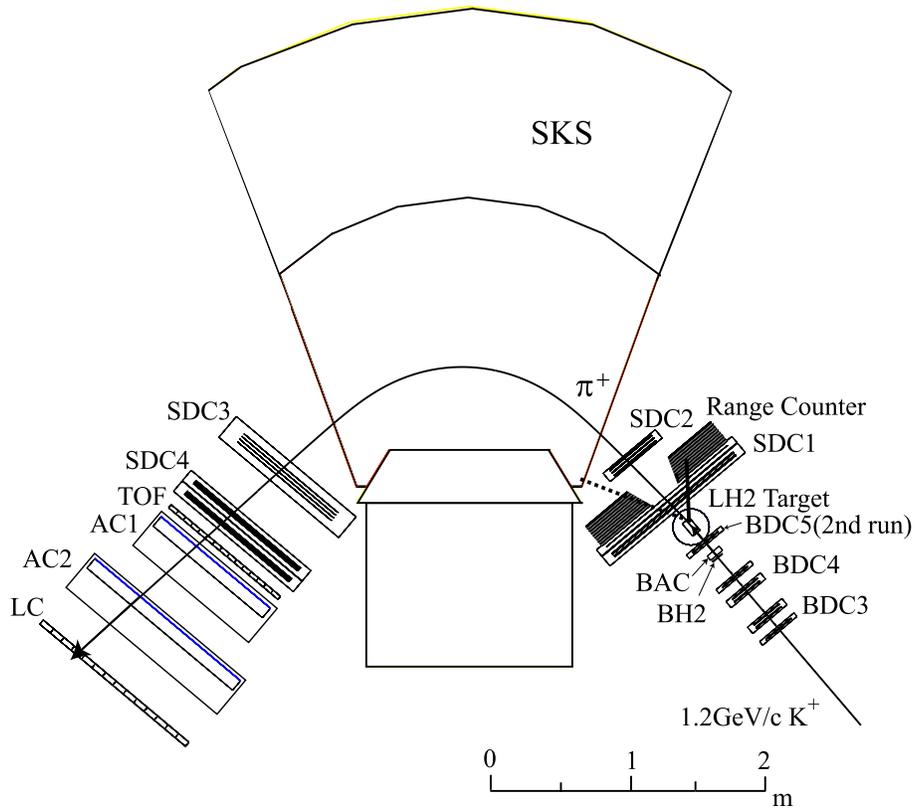}
\caption[]{Experimental setup of the SKS spectrometer.
The SDC1 and a Range Counter (RC) were newly installed in order to
detect the charged particles other than a $\pi^{+}$ detected with SKS.
The BDC5 was used only in the 2nd run to reject reaction events at BAC and BH2.
}
\label{E559SetUp_around_Target}
\end{figure*}

In order to search for the $\Theta^{+}$ via the $K^{+}p \to \pi^{+} \Theta^{+}$
reaction using a high-resolution spectrometer called SKS \cite{SKS},
an experiment has been performed at the K6 beam line
of the KEK 12 GeV proton synchrotron (KEK-PS E559).
%The experimental periods were one month from June 2005 (1st run) and 
%two weeks from December 2005 (2nd run).
Data were taken in two separated periods;
one month in June 2005 (1st run) and two weeks in December 2005 (2nd run).
We searched for the $\Theta^{+}$ using the missing mass technique 
by detecting incident $K^{+}$ beam particles and an outgoing $\pi^{+}$ 
using the K6 beam line spectrometer and the SKS spectrometer, respectively.
Originally, the K6 beam line and SKS spectrometer were constructed
for high-resolution spectroscopy of $\Lambda$ hypernuclei using 
high intensity $\pi^{+}$ beam particles \cite{Hotchi, Hashimoto}.
The advantage of the present experiment
is the excellent mass resolution of 2.4 MeV/$c^{2}$ (FWHM)
which is comparable to the expected width of the $\Theta^{+}$.
In this reaction,
a large background is present, which originates from
$\Delta$ and $K^{*}$ productions with cross section of $\sim$5mb \cite{Bland},
in addition to decays of the incident $K^{+}$ beam.
%If the resolution is poor, the sensitivity becomes quite low.
The SKS provides good mass resolution leading to high sensitivity.
%Therefore this experiment is unique one.
The experimental setups of the K6 beam line and SKS spectrometers
are shown in Fig.~\ref{k6_spectrometer} and \ref{E559SetUp_around_Target}.

In the present experiment, 
for the first time a $K^{+}$ beam was used in this beam line.
The central momentum of the $K^{+}$ beam was set to 1.2 GeV/$c$, 
which was the maximum momentum of the K6 beam line.
The K6 beam line supplied various mass-separated secondary beams
using an electrostatic separator with two correction magnets.
The $K^{+}/\pi^{+}$ ratio in the beam was about 0.1.
The incident particle was defined by two hodoscopes (BH1, BH2)
placed about 9 m apart.
Protons in the beam were rejected at the trigger level 
using a time difference of 7 ns between protons and $K^{+}$'s.
In order to reject beam $\pi^{+}$'s in the online trigger,
an Aerogel Cherenkov counter (BAC), 
whose index of refraction was 1.06, was
installed just upstream of a liquid hydrogen target system.
The efficiency was 98.7$\pm$0.1\% for 1.2 GeV/$c$ $\pi^{+}$'s,
while the overkill rate for $K^{+}$'s was 3.0$\pm$0.2\%.
Each incident particle was analysed with the K6 beam line spectrometer
which consisted of a QQDQQ magnet system and drift chambers 
(BDC1$\cdot$2$\cdot$3$\cdot$4) located upstream and downstream 
of these magnets.  The bending angle was 60 degree.
The expected momentum resolution was $\Delta p/p$=0.047\% (FWHM).

%\begin{figure*} [!t]
%\includegraphics[width=6cm]{E559SetUp_around_Target.eps}
%\caption[]{Experimental setup around the LH$_2$ target.
%The SDC1 and a Range Counter (RC) were newly installed in order to
%detect the charged particles other than a $\pi^{+}$ detected with SKS.
%BDC5 was used only in the 2nd run to reject reaction events at BAC and BH2.
%}
%\label{E559SetUp_around_Target}
%\end{figure*}

\begin{table*}
\caption{Summary of the obtained data}
\label{tab:summary_data_taking}
\begin{ruledtabular}
\begin{tabular}{ccccc}
Reaction& Beam & Target & Number of beam (1st)  & Number of beam (2nd) \\
\hline
($K^{+}, \pi^{+}$) &1.2GeV/$c$  $K^{+}$ & LH$_2$ & $3.31 \times 10^{9}$ & $2.17 \times 10^{9}$\\
\hline
($\pi^{+}, \pi^{+}$) &1.2GeV/$c$  $\pi^{+}$ & LH$_2$ &   &\\
($K^{+}, \pi^{+}$) &1.2GeV/$c$  $K^{+}$ & empty & $2.35 \times 10^{8}$ & $3.87 \times 10^{8}$\\
($\pi^{+}, K^{+}$) &1.1GeV/$c$  $\pi^{+}$ &  LH$_2$ & $8.67 \times 10^{9}$ & $1.71 \times 10^{9}$\\
\end{tabular}
\end{ruledtabular}
\end{table*}

A newly developed liquid hydrogen target had a cooling system
using a liquid helium.
The hydrogen vessel was made from PET (Polyethylene Terephthalate)
for the cylinder, and Mylar for the end caps.
The target was 6.87 cm in diameter and 12.5 cm in length.
%A cylindrical PET, whose diameter and length were 6.87 cm and 12.5 cm,
%was used as a container of a liquid hydrogen.
During the experimental period, the volume and pressure of the target
were monitored and fluctuated less than 1\%.

The scattered $\pi^{+}$'s were detected with the SKS spectrometer located
at forward angles.
The magnetic field was 1.6 T, where the momentum of the central
trajectory was 0.52 GeV/$c$.
The SKS covered the angular region smaller than 25$^\circ$
and had an acceptance in the laboratory
frame of $\sim$0.11 sr for particles whose momentum was 0.46~-~
0.60 GeV/$c$.
The momentum of each outgoing particle was analysed with four drift chambers
(SDC1$\cdot$2$\cdot$3$\cdot$4) located upstream and downstream
of the SKS magnet.
The momentum resolution was estimated to be $\Delta p/p$=0.42\% (FWHM)
from the difference of momenta analysed by both spectrometers
for beam-through events.
For particle identification, trigger counters (TOF, AC1$\cdot$2, LC)
were placed downstream of SDC4.
Both TOF and LC hits were required to select a $\pi^{+}$ in the online
trigger.
Outgoing protons were rejected using LC.
The LC efficiency was typically 95\% for $\pi^{+}$'s.
In offline analysis, the mass was reconstructed from the momentum
and the time-of-flight between BH2 and TOF.

In order to suppress background events,
we modified the detector setup around the target 
as shown in Fig.~\ref{E559SetUp_around_Target}.
For background events, there are hadronic reactions
such as $K^{+} p \to \Delta K \to \pi^{+}KN$ and 
$K^{+} p \to N K^{*} \to \pi^{+}KN$.
The total cross section of each reaction has been measured to be 
3.75$\pm$0.32 mb and 1.06$\pm$0.20 mb, respectively \cite{Bland}.
However, the most serious background came from decay events of $K^{+}$ beam
particles.
Although the kinematical distribution of 
these two-body decays was out of the SKS acceptance,
three-body decays ($K^{+} \to \pi^{+}\pi^{+}\pi^{-}, \pi^{+}\pi^{0}\pi^{0},
\mu^{+}\pi^{0}\nu_{\mu}$ and $e^{+}\pi^{0}\nu_{e}$) could be a large background
which was ten times larger than that from hadronic reactions.
Therefore the rejection of these decay events was crucial in the 
search for the $\Theta^{+}$.
The decay events were separated from the reaction events using 
the difference of the angular distribution and the number of 
charged particles emitted in each event.
For the three-body decays of the $K^{+}$, one or three charged particles 
were emitted in the forward angle within 20$^\circ$,
whereas 
two or four charged particles were emitted in the hadronic reaction events 
with final states $\pi^{+}K^{+}n$ or $\pi^{+}K^{0}p$.
When a $\pi^{+}$ was detected by the SKS, other charged particles were
emitted at large angles (0~-~100 degrees for a $K^{+}$ and
0~-~50 degrees for a proton).
Therefore, in order to detect all possible charged particles and 
measure the angle, a large acceptance drift chamber (SDC1) 
whose effective area was 1200mm$\times$1200mm was installed 
just downstream of the LH$_2$ target.
The SDC1 covered from 0 to 60 degrees.
The SDC1 consisted of five planes (X,X',Y,Y',U).
The maximum drift lengths were 4.5mm and 9.0mm for 
X, X', Y and Y' planes and U plane, respectively.
The SDC1 measured the angle using hit information and vertex position 
obtained from trajectories of the $K^{+}$ and the $\pi^{+}$.
The decay events were suppressed in the offline analysis
using SDC1.
In order to improve the signal-to-noise $S/N$ ratio to detect an 
emitted $K^{+}$,
a range counter (RC) which had the same effective area with
SDC1 was installed just downstream of SDC1.
The RC consisted of 10 layers of plastic scintillators and 9 layers of 
brass absorbers, placed alternately.
The thicknesses of the scintillator and the brass absorber
were 8mm and 9mm, respectively.
Each scintillator layer was segmented horizontally into
5 segments.
The light from the scintillator was collected from PMT's
attached on the top and bottom ends.
In order to avoid the scattering of outgoing $\pi^{+}$'s 
detected with SKS from the material of RC,
there was a hole at the entrance region to the SKS magnet.

In total, $3.31\times 10^{9}$ and $2.17\times 10^{9}$ $K^{+}$
beam particles were irradiated in the 1st and 2nd runs, respectively.
In addition to ($K^{+},\pi^{+}$) events, we took ($\pi^{+}, \pi^{+}$)
events in order to estimate some of cut efficiencies.
The mass scale and resolution were calibrated
using a $\Sigma^{+}$ peak produced via the 
$\pi^{+} p \to K^{+} \Sigma^{+}$ reaction using a 1.1 GeV/$c$ $\pi^{+}$ beams.
The obtained data are summarized in Table~\ref{tab:summary_data_taking}.

\section{Analysis \& Results}

\begin{figure}[h]
\includegraphics[width=8cm]{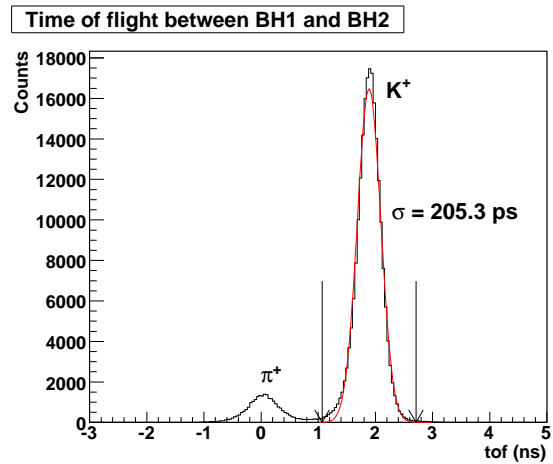}
\caption[]{
Time-of-flight distributions between BH1 and BH2.The arrows indicate the
selection window for $K^{+}$ beam particles which is the $\pm 4\sigma$
region of its time resolution.
}
\label{showBeamCTof2}
\end{figure}

The analysis of beam particles consisted of the identification of $K^{+}$
particles and the analysis of the momentum.
The incident $K^{+}$ beam particles were identified 
with time-of-flight between BH1 and BH2 
as shown in Fig~\ref{showBeamCTof2}.
A typical time difference of the $\pi^{+}$ and $K^{+}$ was 1.9 ns.
The time resolution after a correction using the ADC was $\sigma=205$ ps.
We selected a $\pm 4\sigma$ region to identify $K^{+}$ particles.
%Contamination of $\pi^{+}$ was less than ***\% and negligible.
The beam momentum was reconstructed using hit information of
the drift chambers and a third order transfer matrix.
%The central momentum was 1.21 GeV/$c$ and the momentum spread
%was $\pm 0.04$ GeV/$c$.

\begin{figure}[t]
\includegraphics[width=8cm]{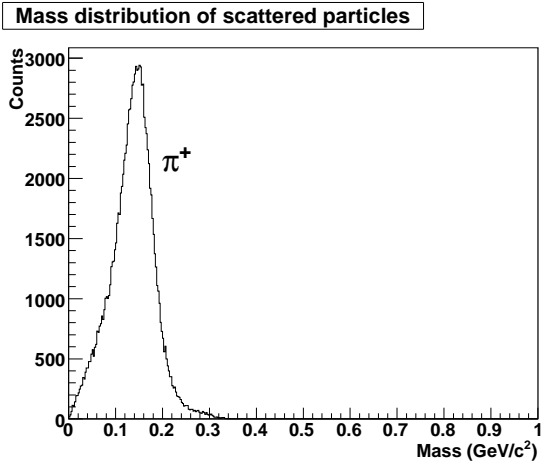}
\caption[]{
Distribution of obtained masses of scattered particles.
}
\label{showScatPart}
\end{figure}

The trajectory of each scattered particle was reconstructed from
its position in the drift chambers and a field map of the SKS magnet.
The straight tracks were defined by fitting 
locally upstream and downstream of the SKS magnet.
These local tracks were connected using the Runge-Kutta method \cite{Runge-Kutta}.
Tracks of reduced $\chi^{2}<100$ were selected as good tracks.
In order to determine the $\chi^2$ limit, 
the peak width of the 
$\Sigma^{+}$ was checked for each reduced $\chi^{2}$ region.
The cut position was determined as a region where a reasonable
resolution (less than 3.3 MeV/$c^{2}$ (FWHM)) was obtained.
From the reconstructed mass distribution, the scattered $\pi^{+}$'s
were clearly selected 
as shown in Fig.~\ref{showScatPart}.

\begin{figure}[t]
\includegraphics[width=8cm]{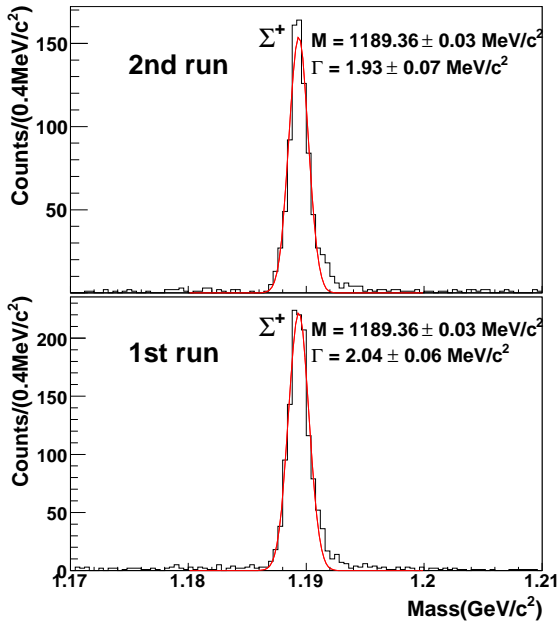}
\caption[]{
Missing mass spectra of the $\pi^{+}p \to K^{+}X$ reaction for
the 1st and 2nd runs.
The peak of $\Sigma^{+}$ is seen.
The obtained widths are consistent with the expected value of
1.98 MeV/$c^{2}$ obtained from a Monte Carlo simulation.
The peak positions for both runs are consistent with each other.
}
\label{showSigma_1st2nd_ver1}
\end{figure}

The validity of the analyses of both spectrometers was confirmed
using the $\pi^{+}p \to K^{+}\Sigma^{+}$ reaction taken for calibration
with a 1.1 GeV/$c$ $\pi^{+}$ beam.
The missing mass spectra of the $\pi^{+} p \to K^{+} X$ reaction
are shown in Fig.~\ref{showSigma_1st2nd_ver1} for the 1st and 2nd runs,
where clear peaks of $\Sigma^{+}$ are identified.
The obtained widths for both runs are consistent with the expected
value of 1.98 MeV/$c^{2}$  from a Monte Carlo simulation
considering the momentum resolution of $\Delta p/p$=0.047\% (FWHM) and 
$\Delta p/p$=0.43\% (FWHM) for the K6 and SKS spectrometers.
The absolute value of the momentum of outgoing particle was 
adjusted to make the $\Sigma^{+}$ peak consistent with its known value.
The mass shift from the known value was 1 MeV/$c^{2}$
which corresponded to 3 MeV/$c$ correction of the momentum.
This momentum correction corresponds to 2 MeV/$c^{2}$ shift
for the peak position of the $\Theta^{+}$.
From Monte Carlo simulations, the missing mass resolution for 
the $\Theta^{+}$ was estimated to be 2.4 MeV/$c^{2}$ (FWHM).
Because the peak positions of the $\Sigma^{+}$ are consistent 
for the 1st and 2nd runs,
the missing mass resolution of the $\Theta^{+}$ is not expected to decrease
by adding the missing mass spectra of the 1st and 2nd runs.

%In order to check the systematics of $(K^{+}, \pi^{+})$ reaction,
%where there is no good hadron resonance as a reference,
%we used a $K^{+} \to \pi^{+}\pi^{+}\pi^{-}$ decay event,
%where both $\pi^{+}$'s were detected with the SKS.
%From the missing mass of $K^{+} \o \pi^{+}\pi^{+}X$ reaction,
%the spectrum of $\pi^{-}$ peak was obtained.
%The obtained width of the $\pi^{-}$ peak was consistent with 
%a simulated value.
%On the other hand, the position was 7.5 MeV/$c^{2}$ higher than
%the PDG value, which means that there is systematic error
%in analyzed momentum of incident or outgoing particle.
%From this result, we estimated the systematic error of the absolute value
%of missing mass for the $\Theta^{+}$ to be 5$\sim$7 MeV/$c^{2}$.

\begin{figure}[t]
\includegraphics[width=8cm]{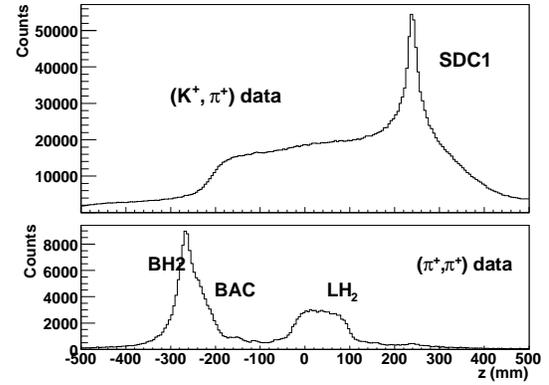}
\caption[]{
Vertex distributions of the ($K^{+}, \pi^{+}$) and ($\pi^{+}, \pi^{+}$) data.
In the vertex distribution of the ($\pi^{+}, \pi^{+}$) data,
the image of the LH$_{2}$ target as well as BH$_2$ and BAC is seen.
The distribution in the ($K^{+}, \pi^{+}$) data was flat from BAC position
($z=-220$) because almost all events of the
($K^{+}, \pi^{+}$) data are decay events.
The peak structure around $z=230$mm was due to miscalculated events 
of $K^{+}$ beam particles, which decayed between SDC1 and SDC2.
}
\label{rgkVtxKpi_Pipi_1st}
\end{figure}

\begin{figure}[h!]
\includegraphics[width=8cm]{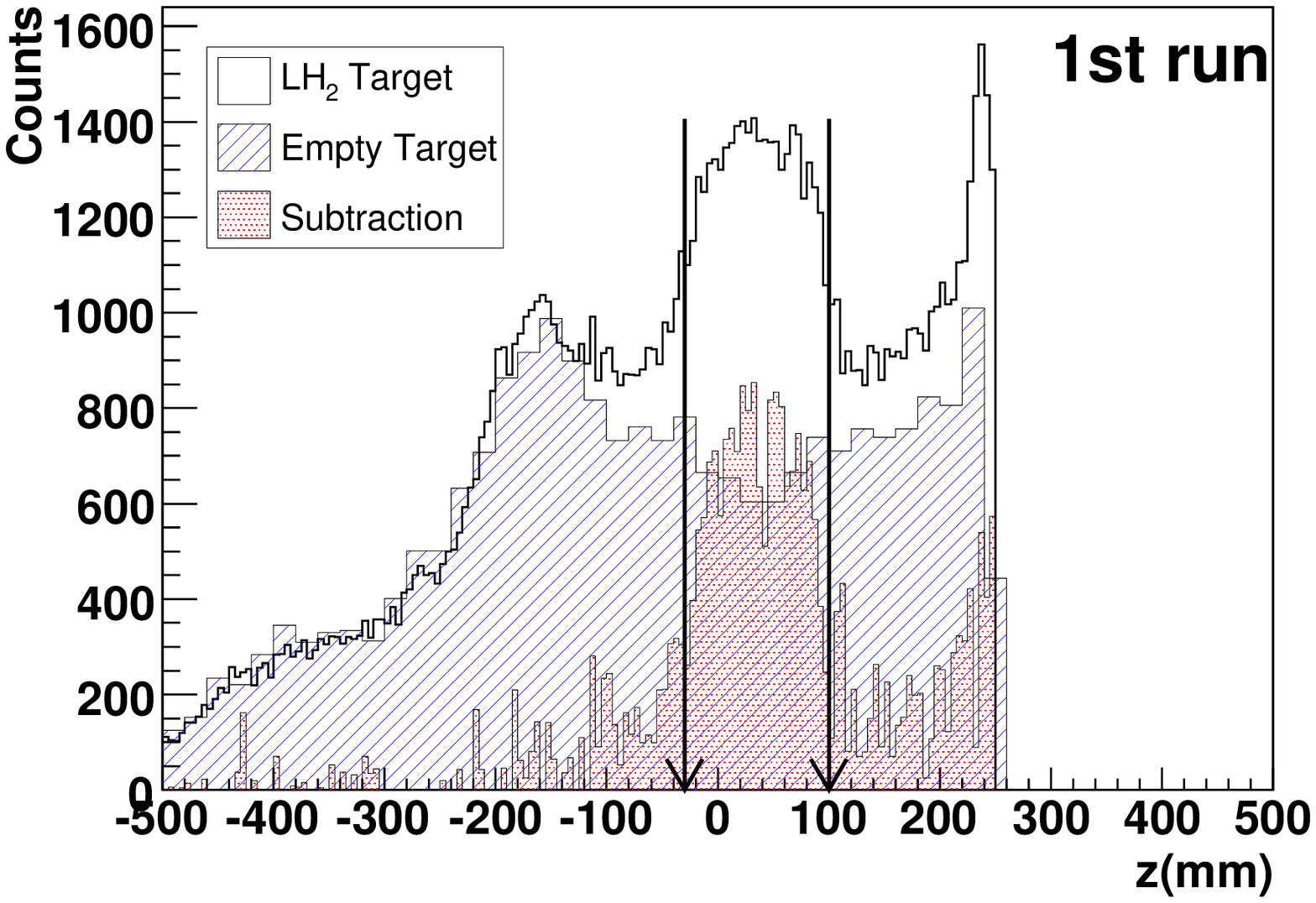}
\includegraphics[width=8cm]{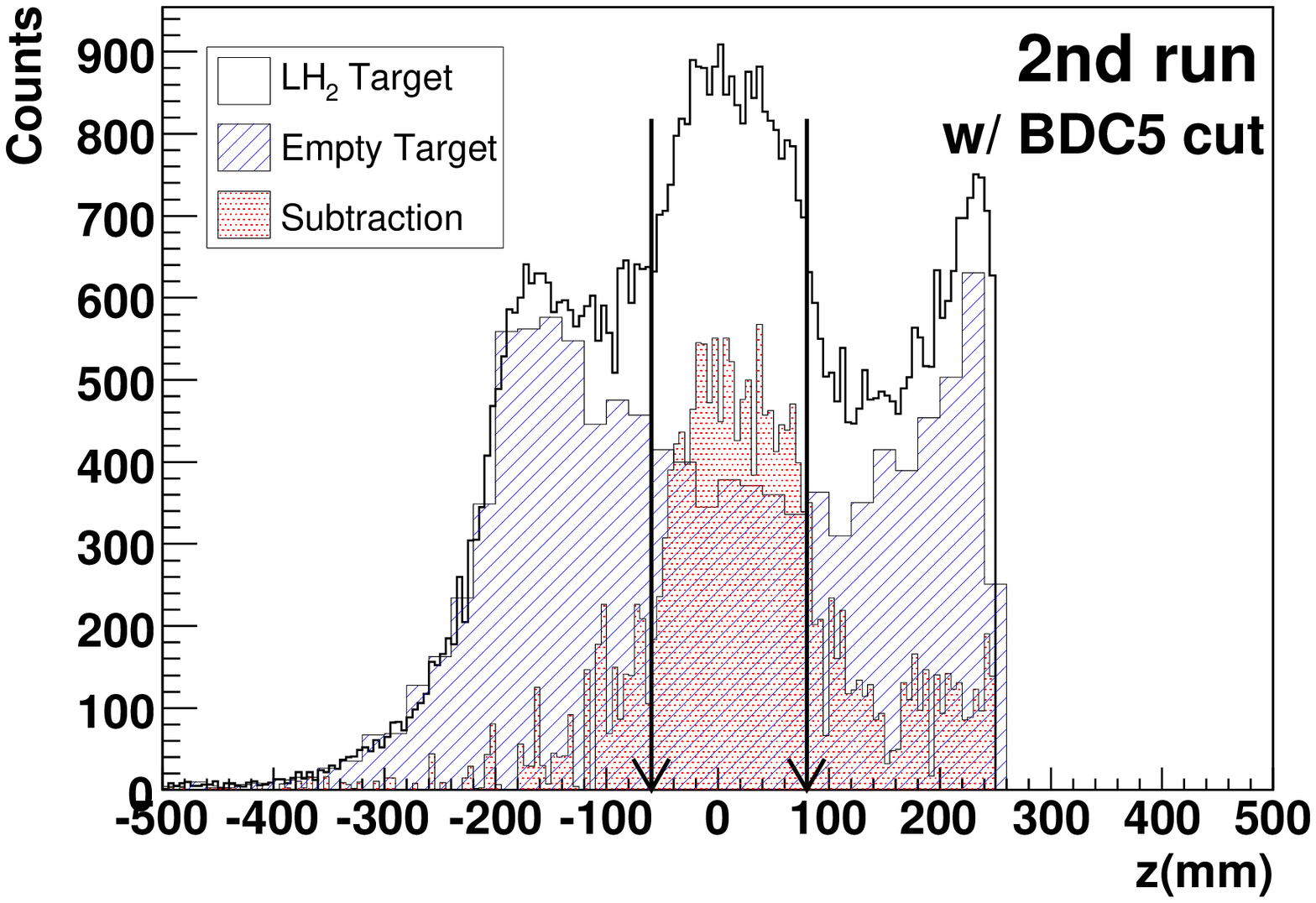}
\caption[]{
Vertex distributions of the ($K^{+}, \pi^{+}$) reaction after the decay 
suppression using SDC1 for the 1st and 2nd  runs.
The open histogram shows the vertex distribution 
obtained using the LH$_2$ target data.
The blue hatched histogram shows the empty target data, which
are normalized using the beam flux.
The red dotted histogram shows the subtraction of these histograms, 
which shows a net contribution of the $K^{+}p \to \pi^{+}X$ reaction events.
The arrows show the cut position for the vertex cut.
}
\label{compVtxWithEmptyAll}
\end{figure}

The vertex position was reconstructed at the point of 
the distance of closest approach 
between the incident and the scattered tracks.
Fig.~\ref{rgkVtxKpi_Pipi_1st} shows the distribution of the 
vertex positions of the 
($K^{+}, \pi^{+}$) and ($\pi^{+}, \pi^{+}$) data.
In the vertex distribution of the ($\pi^{+}, \pi^{+}$) data,
the image of the LH$_{2}$ target, as well as BH2 and BAC, are clearly
seen.
On the other hand, in the ($K^{+}, \pi^{+}$) data, we could not
recognized the LH$_{2}$ target because almost all events of
the ($K^{+}, \pi^{+}$) data are decay events.
The peak structure around $z=230$mm, 
which corresponded to the position of SDC1, was due to 
miscalculated events of $K^{+}$
beam particles that decayed between SDC1 and SDC2.
In order to suppress these decay events,
we used information on the number of tracks
%, which was required to be more than two, 
just after the target and their angles.
In order to detect all possible charged particles,
a large acceptance chamber (SDC1) was installed.
Tracks other than $\pi^{+}$ measured with SKS were 
reconstructed from the hit information of SDC1 and the 
1st layer of RC, and the reconstructed vertex position.
Because more than two charged particles were emitted
at large angles for hadronic reaction,
the number of tracks just after the target
was required to be more than two. 
When the number detected by SDC1 was two, i.e. one more particle
other than $\pi^{+}$ detected with SKS,
the angle of the second particle was required to be larger than 10$^\circ$.
When the number was more than three, the angles of the second and third
particles were required to be more than 15$^\circ$ and 23$^\circ$, respectively.
The cut positions of these angles were determined so as to make
$S/\sqrt{N}$ a maximum, where $S$ is the number of hadronic reaction
events and $N$ is the decay events in the vertex region between
the arrows in Fig.~\ref{compVtxWithEmptyAll}.
Fig.~\ref{compVtxWithEmptyAll} shows the vertex distribution
after this analysis of SDC1.
The vertex of the LH$_{2}$ target was seen once the decay events
were removed.
The hatched histogram shows the empty target data which
is normalized using the beam flux.
The dotted spectrum shows the subtraction of
histograms of the LH$_{2}$ target data and the normalized empty target data.
The subtracted spectrum shows the net contribution 
of the $K^{+}p \to \pi^{+}X$ reaction at the LH$_{2}$ target.
The number of reaction events 
are $1.7\times10^{4}$ and $1.2\times10^{4}$, 
respectively, for the 1st and 2nd runs.

Using the analysis of SDC1, 95\% of decay events at the target region
were removed.
As shown in Fig.~\ref{compVtxWithEmptyAll},
there are still decay events.
One can test whether the amount of decay events is consistent or not
by comparing with a Monte Carlo simulation.
Based on the Monte Carlo simulation 
where efficiencies of the chambers were taken into account,
the suppression factor for the decay events was estimated to be 98\%.
The reason of this inefficiency was that one of the pions 
from three-body decay disappeared by the interaction with the LH$_{2}$
target or materials around the target.
The $e^{+}e^{-}$ from $\pi^{0}$ decay also created mistaken tracks
and decreased the efficiency.
%The contribution due to this 2\% inefficiency is shown by
%the red histogram. 
From this study, we found that there is $\sim$3\% difference 
between the real data and the simulation.
%There are still ununderstandable contribution.
To explain this difference,
we considered whether the $K^{0}_{S}$'s, which were produced at BH2 or BAC
via hadronic reactions such as $K^{+}n \to K^{0}_{S}p$, could contribute
to the vertex distribution, because the $K^{0}_{S}$'s which decayed into 
$\pi^{+}\pi^{-}$  at the target region made a miscalculated vertex
at the cross point between the $\pi^{+}$ and the $K^{+}$ beam.
In the 2nd run, in order to remove these events, 
a new drift chamber (BDC5) with two planes (X, X') was installed 
between the target and BAC.
We applied the following cuts.
At first, the hit position of BDC5 was required to be
consistent with the extrapolated
position from the track defined with BDC3$\cdot$4.
We also required the multiplicity of BDC5 to be one.
The bottom figure in Fig.~\ref{compVtxWithEmptyAll} shows the vertex
distribution applied the BDC5 cut in the 2nd run, 
which shows a slight improvement of the $S/N$ ratio.
However there is still about 2\% inconsistency between the analysis and 
the simulation.
This could be due to the imperfection of the Monte Carlo simulation,
since we could not reproduce the complex LH$_2$ target system
where some materials were not installed.
Another possible reason is incompleteness of the analysis of SDC1.
We took this 2\% inconsistency into account as the systematic error of 
the efficiency of the SDC1 analysis described in the following paragraph.

\begin{table*} [!t]
\begin{ruledtabular}
\caption{Summary of the cuts and their efficiencies.}
\label{tab:summary_ana_eff}
\begin{tabular}{ccc}
& cut & efficiency(\%) (1st run / 2nd run)\\
\hline
$f_{K^{+}beam}$ & $K^{+}$ beam on-target factor  & 84.5$\pm$3.5 / 96.2$\pm$1.6 \\
$f_{K^{+}decay}$ & $K^{+}$ beam decay factor  & 96.7$\pm$0.1 / 95.7$\pm$0.1 \\
$\epsilon_{K6}$ & tracking efficiency of beam particle & 95.3$\pm$0.3 / 95.6$\pm$0.3 \\
$\epsilon_{LC}$ & LC efficiency  & 95.5$\pm$0.2 \\
$\epsilon_{TOF}$ & TOF efficiency  & $\sim$100. \\
$f_{\pi^{+}decay}$ & $\pi^{+}$ decay factor & 85.2$\pm$0.2 \\
$f_{\pi^{+} int}$ & $\pi^{+}$ interaction factor  & 94$\pm$2 \\
$\epsilon_{SdcIn}$ & SdcIn tracking efficiency  & 87 $\pm$1 \\
$\epsilon_{SdcOut}$ & SdcOut tracking efficiency  & 92.9$\pm$0.3 \\
$\epsilon_{Sks}$ & Sks tracking efficiency  & 95$\pm$0.7 \\
$\epsilon_{Sdc1}$ & Sdc1 analysis efficiency  & 69.4$\pm$3.7 \\
$\epsilon_{vtx}$ & vertex cut efficiency  & 85.2$^{+2.9}_{-1.3}$ / 85.0$^{+0.4}_{-0.9}$ \\
$\epsilon_{Bdc5}$ & BDC5 cut efficiency  & ------ / 91.6$\pm$0.2 \\
\hline
$d\Omega$ & acceptance at lab.~frame  & 0.11sr. \\
\end{tabular}
\end{ruledtabular}
\end{table*}

The efficiency of the SDC1 analysis 
for the $\Theta^{+}$ events was also checked
by this Monte Carlo simulation.
Because charged particles other than $\pi^{+}$ in SKS
were required to be detected by SDC1,
the acceptance of the chamber must be considered for these 
particles and the tracking efficiency of the SDC1 analysis.
In this simulation, we assumed the following three kinds of 
angular distributions
of $\pi^{+}$ in the center of mass system;
\begin{enumerate}
\item flat distribution, 
\item forward peak distribution ((1$+\cos\theta$)/2),
\item backward peak distribution ((1$-\cos\theta$)/2),
\end{enumerate}
where $\theta$ is the angle of $\pi^{+}$ in the $K^{+}p$ center of mass 
(c.m.) system.
For the decay distribution of $\Theta^{+}\to KN$, 
a flat distribution was assumed.
The branching ratios of $\Theta^{+} \to K^{+}n, K^{0}_{S}p$
and $K^{0}_{L}p$ were assumed to be 50\%, 25\% and 25\%, respectively.
Between these distributions, the difference was small and
the efficiency was estimated to be 69.4$\pm$3.7\%.
The error consisted of two parts.
The first one is the deviation 
between the different angular distributions.
The other one is the 2\% uncertainty of tracking analysis using SDC1
as described in the previous paragraph.
%The deviation between the assumed angular distribution gives a 
%2\% uncertainty of tracking analysis using the SDC1
%(as described in previous paragraph).

\begin{figure}[t]
\includegraphics[width=8cm]{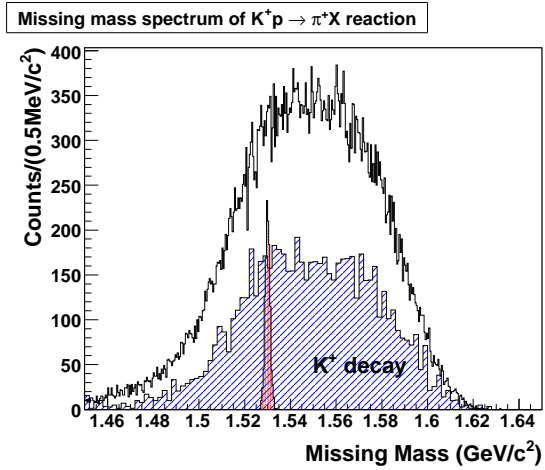}
\caption[]{
Missing mass spectrum of the $K^{+}p \to \pi^{+}X$ reaction
where the spectra of the 1st and 2nd runs are added.
The hatched histogram shows the empty target data, which are normalized 
using the beam flux.
The dotted spectrum shows the expected spectrum assuming that
the width of the $\Theta^{+}$ is 0 MeV/$c^{2}$, the total cross section of the $K^{+}p \to \pi^{+}\Theta^{+}$ reaction
is 50 $\mu$b and the angular distribution of the $\pi^{+}$ is isotropic 
in the $K^{+}p$ c.m. system.
}
\label{missmass}
\end{figure}

Fig.~\ref{missmass} shows the missing mass spectrum of the 
$K^{+}p \to \pi^{+}X$ reaction after selecting the LH$_{2}$ 
target regions using the vertex distribution.
In this figure, the data of the 1st and 2nd runs are added.
The hatched histogram shows the empty target data, which are normalized 
using the beam flux.
This spectrum shows the contribution from the $K^{+}$ decays
in the region between the arrows in Fig.~\ref{compVtxWithEmptyAll}.
The dotted spectrum shows the expected spectrum assuming that
the width of the $\Theta^{+}$ is 0 MeV/$c^{2}$, 
the total cross section of the $K^{+}p \to \pi^{+}\Theta^{+}$ reaction
is 50 $\mu$b and the angular distribution of the $\pi^{+}$ is isotropic 
in the $K^{+}p$ c.m. system.
In the present experiment, no significant peak was observed.
We also obtained the differential cross section averaged over 
2$^{\rm o}$ to 22$^{\rm o}$ in the laboratory frame.
The differential cross section is defined by the following equation,

%\begin{widetext}
%\begin{equation}
\begin{eqnarray}
  \left(
  \frac{d\sigma}{d\Omega}
  \right)
  & = & \frac{1}{N_{target}}\cdot \frac{1}{N_{beam}\cdot f_{K^{+}beam} \cdot f_{K^{+}decay} \cdot \epsilon_{K6}} \nonumber \\
  &   & \cdot \frac{1}{\epsilon_{LC} \cdot \epsilon_{TOF}} \cdot \frac{1}{f_{\pi^{+}decay} \cdot f_{\pi^{+} int}} \nonumber \\
  &   & \cdot \frac{N_{\Theta^{+}}} {\epsilon_{SdcIn} \cdot \epsilon_{SdcOut} \cdot \epsilon_{Sks} \cdot \epsilon_{Sdc1}\cdot \epsilon_{vtx} \cdot d\Omega}, \label{eq:cross_section}
\end{eqnarray}
%\end{equation}
%\end{widetext}
where the $f$'s and $\epsilon$'s represent the correction factors and 
efficiencies whose typical values are summarized 
in Table~\ref{tab:summary_ana_eff}.

The $f_{K^{+}beam}$ represents the ratio of the number of $K^{+}$'s 
which hit the LH$_{2}$ target to that of $K^{+}$'s identified from 
the time-of-flight between BH1 and BH2.
Because the horizontal size of the beam at the target region
was comparable with the target size,
part of the beam particles did not pass through the target.
The $f_{K^{+}beam}$'s are estimated to be 84.5 $\pm$ 3.5\% and 
96.2 $\pm$ 1.6\% for the 1st and 2nd runs, respectively, 
using data taken with a KBEAM trigger
where there was no bias of scattered particles.
Because the beam center was slightly shifted from the target center
in the 1st run,
the $f_{K^{+}beam}$ in the 1st run was smaller than that in the 2nd run.
The coefficient, $f_{K^{+}decay}$, represents the correction factor
due to the decay in flight of the $K^{+}$ beam particles between
BH2 and the LH$_{2}$ target.
The $f_{K^{+}decay}$'s were estimated to be 96.7 $\pm$ 0.1\% and 
95.7 $\pm$ 0.1\% for the 1st and 2nd runs, respectively.
In the 2nd run, BH2 was moved to upstream in order to install BDC5.
Therefore the $f_{K^{+}decay}$ of the 2nd run was smaller than that of 
the 1st run.

The $\epsilon_{K6}$ represents the tracking efficiency for the incident 
particles,
estimated by the ratio of the number of
events accepted as a good trajectory to that of good beam particles
defined using the time-of-flight between BH1 and BH2
and the energy deposit at BH2.
The $\epsilon_{K6}$'s were obtained to be 95.3$\pm$0.3 and 95.6$\pm$0.3 
for the 1st and 2nd runs, respectively.

The $\epsilon_{LC}$ and $\epsilon_{TOF}$ represent the efficiencies of
LC and TOF counters, respectively.
The LC and TOF were segmented horizontally into 
14 and 15 segments, respectively.
These efficiencies are estimated for each segment using data
taken in the trigger condition without LC and TOF.
The typical value of $\epsilon_{LC}$ was 95.5$\pm$0.2\%.
In order to obtain the cross section, the efficiency of the segment 
through which the outgoing $\pi^{+}$ passed was used.
The typical value of the $\epsilon_{TOF}$ was almost 100\%.

The coefficient, $f_{\pi^{+}decay}$, represents the correction factor
due to the decay-in-flight of the $\pi^{+}$.
In the offline analysis, we required the hits of the LC and TOF
corresponding to the trajectory obtained by the tracking routine
using SDC3$\cdot$4.
From Monte Carlo simulations, we found that events where
$\pi^{+}$'s decayed after SDC4 could be analyzed as good events,
because the angle of the $\mu^{+}$ from the $\pi^{+}$ decay was
less than 4$^\circ$.
Therefore the $f_{\pi^{+}decay}$ was calculated event by event
using the flight length from the vertex point to the exit of SDC4.
The typical value of $f_{\pi^{+}decay}$ was 85.2$\pm$0.2\%.
The coefficient, $f_{\pi^{+}int}$, represents the correction factor
due to the interaction rate of $\pi^{+}$ in the materials of the target
and the SKS spectrometer.
The factor was calculated with the Monte Carlo simulation based on
GEANT4.
The value of $f_{\pi^{+}int}$ was found to be 94$\pm$2\%.

The coefficients, $\epsilon_{SdcIn}$ and $\epsilon_{SdcOut}$, 
represent the efficiencies of the local tracking 
upstream and downstream of the SKS magnet.
The $\epsilon_{SdcIn}$ was estimated from Monte Carlo simulations
to be 87$\pm$1\%.
Typically there were multiple hits in SDC1 because it was designed
to detect many charged particles at once.
Therefore hits of SDC1 and SDC2
which did not originate from a single particle were connected
in the tracking routine and finally rejected by the further analysis.
The $\epsilon_{SdcOut}$ was estimated using the data of five 
beam-through runs where the $\pi^{+}$ beams of fixed momentum from
0.475 to 0.525 GeV/$c$ were directly analyzed with the SKS spectrometer.
The efficiency was constant for all horizontal positions of 
the outgoing particle.
The $\epsilon_{SdcOut}$ was obtained to be 92.9$\pm$0.3\%.
The coefficient, $\epsilon_{SKS}$, represents the efficiency of the 
Runge-Kutta tracking which calculates the trajectory of 
the outgoing particle by connecting local tracks.
The $\epsilon_{SKS}$ was estimated by using 
scattered protons selected only using TOF and LC.
%In this case, the particle decay in flight was negligible.
The efficiency depended on the slope in the vertical plane (dy/dz) of the 
outgoing particle at the target.
For small slope (dy/dz $\sim$ 0), the $\epsilon_{SKS}$ was 95$\pm$0.7\%.
For large slope (dy/dz $\sim$ $\pm0.8$), the $\epsilon_{SKS}$ was 89$\pm$1\%.
The efficiency was corrected according to the 
slope of each trajectory.

The coefficient, $\epsilon_{Sdc1}$, represents the efficiency of the SDC1 
analysis described in the previous paragraph.
The $\epsilon_{Sdc1}$ was obtained to be 69.4$\pm$3.7\%.

The coefficient, $\epsilon_{vtx}$, represents the efficiency of the 
vertex cut.
In order to estimate this value, we used the vertex distribution 
of the ($\pi^{+}, \pi^{+}$) events,
because the target image could be identified more precisely.
The $\epsilon_{vtx}$ values are 85.2$^{+2.9}_{-1.0}$\%
and 85.0$^{+0.4}_{-0.9}$\% for the 1st and 2nd runs, respectively.
The coefficient $\epsilon_{Bdc5}$ represents the efficiency of the 
BDC5 analysis applied only in the 2nd run.
This value was also estimated by using the ($\pi^{+}, \pi^{+}$) data
from the ratio of the number of events within the target region
in the vertex distribution with and without the BDC5 cut.
The value of $\epsilon_{Bdc5}$ is 91.6$\pm$0.2\%.

\begin{figure}[t]
\includegraphics[width=8cm]{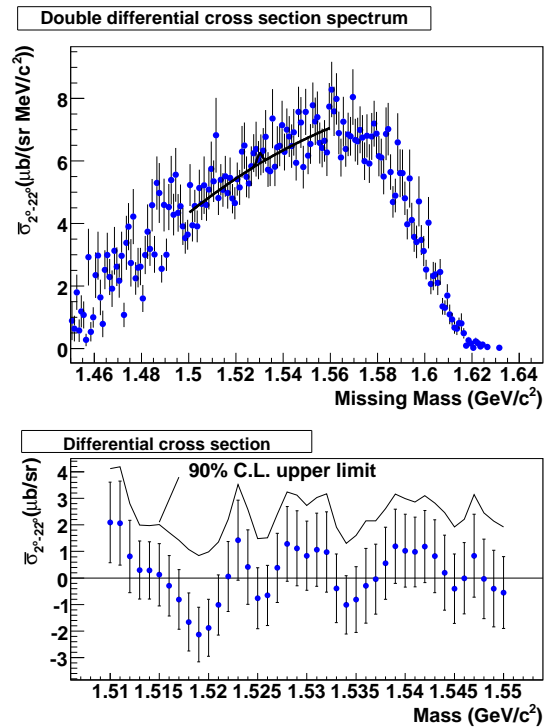}
\caption[]{
Top figure: Double differential cross section of the $K^{+}p \to \pi^{+}X$
reaction averaged over 2$^{\rm o}$ to 22$^{\rm o}$ in the laboratory
frame. 
This spectrum includes a contribution from the $K^{+}$ decay events
which exist between the arrows in Fig.~\ref{compVtxWithEmptyAll}.
In order to derive the upper limit of the differential cross section
of the $K^{+}p \to \pi^{+} \Theta^{+}$ reaction,
we fitted this spectrum with a background using a 2nd order polynomial
function and a Gaussian peak with a fixed width of 2.4 MeV/$c^{2}$ (FWHM).\\
Bottom figure: The upper limits of the differential cross section of the
$K^{+}p \to \pi^{+} \Theta^{+}$ reaction averaged over 
2$^{\rm o}$ to 22$^{\rm o}$ in the laboratory frame as a function of the
mass of the $\Theta^{+}$.
The data points show the value and the error of the
differential cross section obtained from the area of the fitted 
Gaussian function.
The line shows the 90\% C.L. upper limit of the differential cross section.
}
\label{showMassSearch}
\end{figure}

Because some of efficiencies depended on the trajectory of the particle,
%the differential cross section was calculated event by event.
these efficiencies were calculated event by event.
The differential cross section averaged over 
2$^{\rm o}$ to 22$^{\rm o}$ in the laboratory frame was calculated 
using the following equation,

\begin{equation}
\bar{\sigma}_{2^{\rm o}-22^{\rm o}} = \int_{2^{\rm o}}^{22^{\rm o}}
  \left(
  \frac{d\sigma}{d\Omega}
  \right) d\Omega
  /\int_{2^{\rm o}}^{22^{\rm o}} d\Omega .
\end{equation}

Fig.~\ref{showMassSearch} shows the differential cross section 
versus missing mass
which also shows no peak structure.
This spectrum includes the contribution of $K^{+}$ decay events
which exist between the arrows in Fig.~\ref{compVtxWithEmptyAll}.
The error includes both of the statistical and systematic errors.
We derived a 90 \% C.L. upper limit of the differential cross section
of the $K^{+}p \to \pi^{+}\Theta^{+}$ reaction.
As shown in the top figure of Fig.~\ref{showMassSearch},
we fitted this spectrum with a background using a 2nd order polynomial 
function and a Gaussian peak with a width of 2.4 MeV/$c^{2}$ (FWHM)
which is the expected resolution for the $\Theta^{+}$.
The differential cross section was calculated from 
the area of the Gaussian function.
The bottom figure of Fig.~\ref{showMassSearch} shows 
the values and the errors of the 
differential cross section as a function of the peak position.
The solid line in the bottom figure of Fig.~\ref{showMassSearch}
shows the 90 \% C.L. upper limit of the differential cross section
considering that this distribution is based on Gaussian statistics.
%The 90 \% C.L. upper limit of the differential 
%cross section averaged over 2$^{\rm o}$ to 22$^{\rm o}$ in the laboratory frame
This 90 \% C.L. upper limit  
is less than 3.5 $\mu$b/sr for almost all mass region.

%The total production cross section can be estimated 
%when we assume the angular distribution of the $\pi^{+}$
%generated in the $K^{+}p \to \pi^{+}\Theta^{+}$ reaction.
%We assumed three distributions ,i.e., a flat distribution, 
%a forward peak distribution and a backward peak distribution,
%mentioned above.
%The acceptance of the SKS is 2.8\%, 5.5\% and 0.15\% for each distribution.
%The 90\% C.L. uper limit of the total cross section was obtained 
%to be 7$\mu b$, 3.5$\mu b$ and 133$\mu b$, respectively.
%The total cross section depends strongly on the angular distribution of 
%the $\pi^{+}$.
%If the distribution is a backward peak one, our experiment
%does not have an enough sensitivity.

\section{Discussion}

The experimental results are now compared with theoretical calculations.
A production mechanism based on both the present
results of the $K^{+}p \to \pi^{+}\Theta^{+}$ reaction and 
the results of the $\pi^{-}p \to K^{-}\Theta^{+}$ reaction 
will be discussed.

\begin{figure} [b]
\includegraphics[width=7cm]{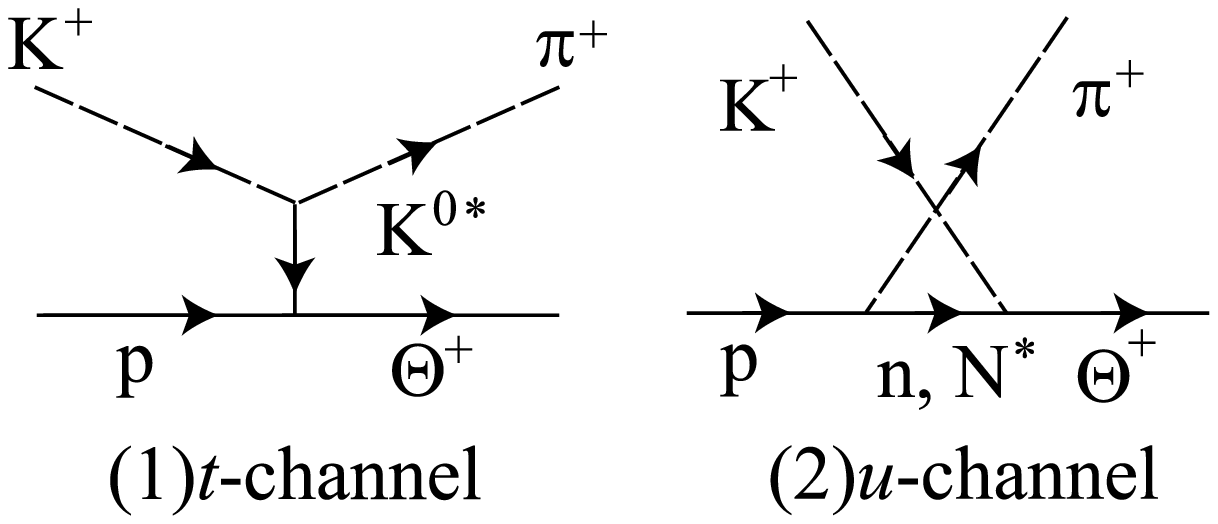}
\caption[]{Diagrams for the $K^{+}p \to \pi^{+}\Theta^{+}$ reaction.}
\label{diagram2}
%\end{figure}
%\begin{figure} [t]
\includegraphics[width=7cm]{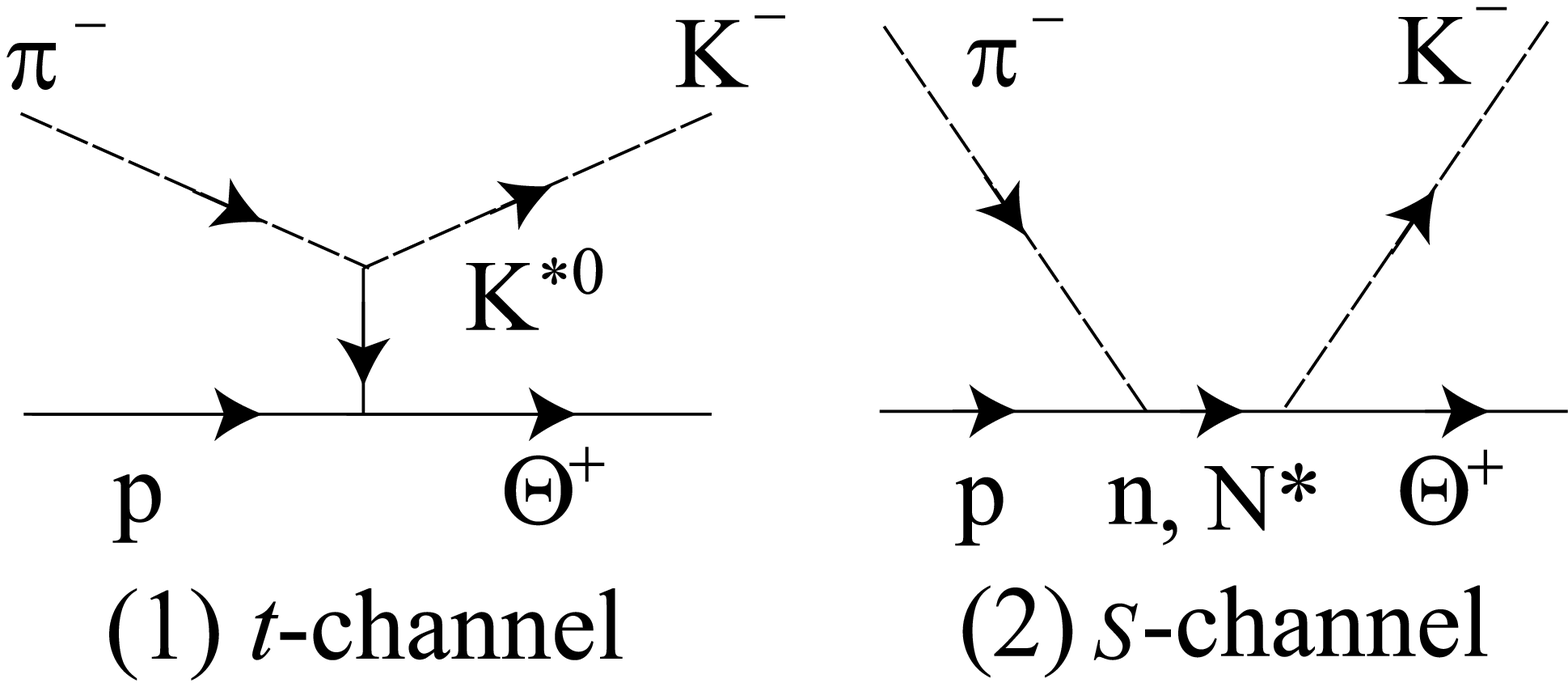}
\caption[]{Diagrams for the $\pi^{-}p \to K^{-}\Theta^{+}$ reaction.}
\label{diagram_pik}
\end{figure}

\begin{figure} [t]
\includegraphics[width=8cm]{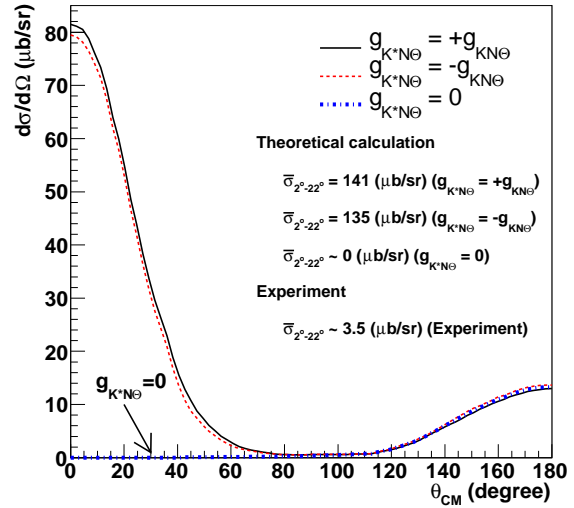}
\caption[]{Differential cross section calculated by Y.~Oh {\it et al.} at 
$\sqrt{s}$=2.4 GeV (the present experiment was carried out at 
$\sqrt{s}$=1.9 GeV). This is the same as Fig.~3 (c) in Ref.~\cite{Oh2}.
The horizontal axis represents the 
scattering angle of the $\pi^{+}$ in the c.m. frame of $K^{+}p$ system.
The solid line is obtained with $g_{K*N\Theta}=+g_{KN\Theta}$,
the dashed line with $g_{K*N\Theta}=-g_{KN\Theta}$,
and the dot-dashed line with $g_{K*N\Theta}=0$.
The differential cross sections averaged over 2$^\circ$ to 22$^\circ$
in the laboratory frame are also listed for each case.}
\label{showTheoryOh}
\end{figure}

In the $K^{+}p\to \pi^{+}\Theta^{+}$ reaction, 
it is possible to consider a $t$-channel process where
$K^{0*}$ is exchanged and a $u$-channel process where $N^{*}$ is 
an intermediate state as shown in Fig.~\ref{diagram2}.
Fig.~\ref{showTheoryOh} shows the differential cross section calculated
by Y.~Oh {\it et al.} using an effective interaction Lagrangian \cite{Oh2}.
Their calculation is controlled by two coupling constants, $g_{KN\Theta}$
and $g_{K^{*}N\Theta}$.
The coupling constant $g_{KN\Theta}$ is related to the
decay width of the $\Theta^{+}$.
They assumed that $g_{KN\Theta}$ was 1.0 which corresponds
to a decay width of 1.03 MeV/$c^{2}$.
On the other hand, there is no experimental information 
about $g_{K^{*}N\Theta}$. 
Therefore they calculated in three cases where $g_{K^{*}N\Theta}=0$,
$g_{K^{*}N\Theta}=g_{KN\Theta}$ and $g_{K^{*}N\Theta}=-g_{KN\Theta}$. 
If there is the $t$-channel process ($g_{K^{*}N\Theta}= \pm g_{KN\Theta}$),
the differential cross section would have 
a forward peak distribution as shown by the
solid and dashed lines in Fig.~\ref{showTheoryOh}.
The calculated differential cross section averaged over 2$^\circ$ to 
22$^\circ$ in the laboratory frame ($\bar{\sigma}_{2^\circ-22^\circ}$)
is about 140 $\mu b$/sr in this case.
The experimental upper limit of 3.5 $\mu$b/sr
is much smaller than this theoretical value.
Therefore the $t$-channel process is excluded by the present results.
On the other hand, if the $t$-channel process does not exist
and only the $u$-channel process exits,
the differential cross section shows a backward peak distribution 
as shown by the dot-dashed line in Fig.~\ref{showTheoryOh}.
In this case, the $\bar{\sigma}_{2^\circ-22^\circ}$ is almost 0 $\mu b$/sr.
Our experiment is not sensitive enough to exclude the $u$-channel process.

The E522 collaboration reported an upper limit of the cross section
of the $\pi^{-}p \to K^{-}\Theta^{+}$ reaction of 3.9 $\mu$b at the 90\% 
confidence level assuming that the $\Theta^{+}$ is produced isotropically 
in the center of mass system.
In the $\pi^{-}p \to K^{-}\Theta^{+}$ reaction, the $t$-channel and 
$s$-channel processes are considered as shown in Fig.~\ref{diagram_pik}.
The cross section is controlled by 
the same coupling constants, $g_{K^{*}N\Theta}$ and $g_{KN\Theta}$, 
which are used for the $K^{+}p \to \pi^{+}\Theta^{+}$ reaction.
In order to explain the small cross section,
the following two things are possible:
1) The coupling constant $g_{KN\Theta}$ is small;
2) Although the coupling constants $g_{K^{*}N\Theta}$ and $g_{KN\Theta}$
are sizable, 
the total cross section becomes small due to
a destructive interference between two amplitudes 
of $g_{K^{*}N\Theta}$ and $g_{KN\Theta}$.
From the result of the $K^{+}p \to \pi^{+}\Theta^{+}$ reaction,
$g_{K^{*}N\Theta}$ is considered to be quite small.
Therefore it is unlikely that the cross section
is small due to the interference.
If $g_{KN\Theta}$ is small, then 
the width of $\Theta^{+}$ is quite narrow.

\begin{figure} [t]
\includegraphics[width=7cm]{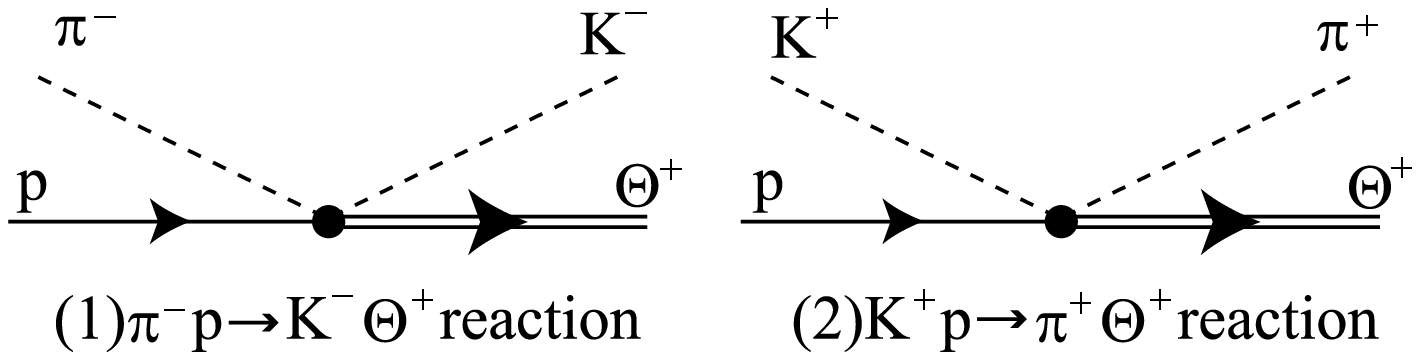}
\caption[]{Diagrams with two-meson coupling for the $\pi^{-}p \to K^{-}\Theta^{+}$ and $K^{+}p \to \pi^{+}\Theta^{+}$ reactions.}
\label{diagram_hyodo}
\end{figure}

Hyodo and Hosaka studied the production mechanism
taking into account the importance of two-meson couplings \cite{Hyodo2}.
They calculated the production cross section of 
the $\pi^{-} p \to K^{-} \Theta^{+}$ and 
$K^{+} p \to \pi^{+} \Theta^{+}$ reactions 
for both $J^{P} = 1/2^{+}$ and $3/2^{-}$ 
using the Feynman diagrams shown in Fig.~\ref{diagram_hyodo}.
They obtained the scalar and vector coupling constants of 
$\Theta K \pi N$, $g^{s}$ and $g^{v}$, using flavor SU(3)
symmetry and the decay width of the $N^{*}(1710) \to \pi\pi N$.
Without a two-meson coupling, all of the amplitudes for the $\Theta^{+}$ 
production are proportional to the $g_{KN\Theta}$ coupling,
which is fixed by the supposed small decay width of the $\Theta^{+}$.
However, even with an extremely narrow width of the $\Theta^{+}$,
a sizable cross section can be obtained using the two-meson coupling
determined from the decay width of $N^{*}(1710)$.
These coupling constants have uncertainty due to
the experimental uncertainties in the branching ratio.
They restricted the coupling constants to be consistent with 
the upper limit of the cross section of the $\pi^{-}p \to K^{-}\Theta^{+}$
reaction.
Moreover the relative phase between scalar and vector coupling constants
could not be determined solely from the decay width of the $N^{*}(1710)$.
This relative phase is quite important because it determines the
interference term of these two amplitudes.
If $g_{s}$ and $g_{v}$ have the same phase,
the two amplitudes interfere constructively for the 
$\pi^{-}p \to K^{-} \Theta^{+}$ channel, 
while in the $K^{+}p \to \pi^{+}\Theta^{+}$ case it gives 
destructive interference.
On the other hand, if $g_{s}$ and $g_{v}$ have the opposite phase,
the situations for constructive and destructive interference reverse.
Considering the small cross section obtained 
in the $\pi^{-}p \to K^{-} \Theta^{+}$ reaction,
these data suggest the latter case.
Then, the cross section of the $K^{+}p \to \pi^{+}\Theta^{+}$
reaction could be large.
They calculated total cross sections of
2.5 mb and 110 $\mu$b in the cases of $J^{P}=1/2^{+}$ and $3/2^{-}$,
respectively.
The $\bar{\sigma}_{2^\circ-22^\circ}$'s 
are $\sim$600 $\mu$b/sr and $\sim$50 $\mu$b/sr in each case.
The experimental upper limit of 3.5 $\mu$b/sr is much smaller
than these calculations.
If they take the same phase for $g_{s}$ and $g_{v}$
in order to explain the cross section of the $K^{+}p \to \pi^{+}\Theta^{+}$
reaction, the cross section of the $\pi^{-}p \to K^{-}\Theta^{+}$
becomes large due to the constructive interference, which is inconsistent
with the experimental result.
Therefore this model can not explain both of the experimental results
simultaneously.
%the experimental result can not be explained by
%this model.
%Therefore the contribution of the two meson coupling 
%is considered negligible.

In summary, we have searched for the $\Theta^{+}$ via 
the $K^{+}p \to  \pi^{+}X$ reaction using
a 1.2 GeV/$c$ $K^{+}$ beam at the K6 beam line of 
the KEK-PS 12 GeV Proton Synchrotron.
In the missing mass spectrum of the $K^{+}p \to \pi^{+}X$ reaction,
no clear peak structure was observed.
A 90 \% C.L. upper limit of the differential cross section, 
averaged over 2$^\circ$ to 22$^\circ$ in the laboratory frame
of the $K^{+}p \to \pi^{+}\Theta^{+}$ reaction, is obtained at 3.5 $\mu$b/sr.
From the present experiment and the experiment by the E522 collaboration,
it is found that both of the production cross sections from 
the $\pi^{-}p \to K^{-}\Theta^{+}$  and
$K^{+}p \to \pi^{+}\Theta^{+}$  reactions are small.
From the small differential cross section at the forward angles
of the $K^{+}p \to \pi^{+}\Theta^{+}$  reaction,
the $t$-channel process, where a $K^{0*}$ is exchanged, is excluded.
Therefore the small cross section of the $\pi^{-}p \to K^{-}\Theta^{+}$
reaction cannot be explained by the destructive interference between
two amplitudes due to the couplings $g_{KN\Theta}$ and $g_{K^{*}N\Theta}$.
In order to explain the small cross section, the coupling constant
$g_{KN\Theta}$ must be small.
%The contribution of the two meson coupling is also 
%considered to be negligibly small.
The model by Hyodo and Hosaka which explains the result of 
the $\pi^{-}p \to K^{-}\Theta^{+}$ reaction also cannot explain
the present result.

\section{Acknowledgments}
We would like to express our thanks to staffs of KEK PS
and beam channel group for their
support to provide beam with the excellent condition during the experiment.
We would like to thanks to RIKEN Radiation Laboratory for the usage of 
RIKEN-CCJ computer system.
Some of the authors (K. M.) and (S. D.) thank to the Japan Society for 
the Promotion of Science (JSPS) for support.
One author (K.H.) thanks the National Science Foundation for support.
This work was supported by the Grant-in-Aid for the 21st Century
COE ''Center for Diversity and Universality in Physics'' from
the Ministry of Ministry of Education, Culture, Science and Technology
(MEXT) of Japan.
This work was supported by the Grant-in-Aid for Specially Promoted Research 
(No.15001001) from the
Ministry of Education, Culture, Science and Technology, Japan.

\end{document}